\documentclass{emulateapj}
\usepackage{psfig}
\usepackage{apjfonts}

\newcommand\Res{\ifmmode{\cal R}\else{${\cal R}$}\fi}
\newcommand\Ha{${\rm H}\alpha$}
\newcommand\Lya{${\rm Ly}\alpha$}
\newcommand\eg{{\it e.g.,}}
\newcommand\ie{{\it i.e.,}}
\newcommand\cf{{\it cf.}}
\newcommand\etal{{\it\ et al.}}
\newcommand\kms{\ifmmode{\rm\ km\ s^{-1}}\else{$\rm\ km\ s^{-1}$}\fi}

\begin{document}

\title{In Search of the Dark Ages -- An Experimental Challenge}

\author{J. Bland-Hawthorn}
\affil{Anglo-Australian Observatory, P.O. Box 296, Epping, NSW 2121, Australia}
\and
\author{P.E.J. Nulsen\altaffilmark{1}}
\affil{Harvard-Smithsonian Center for Astrophysics, 60 Garden Street, Cambridge, MA 02138}
\altaffiltext{1}{On leave from the University of Wollongong}

\begin{abstract}

Most direct source detections beyond $z\sim7$ are likely to arise from
wide-field narrowband surveys of \Lya\ emission in the J band. For
this to be true, the \Lya\ emission must somehow escape from compact 
star-forming regions (CSR) presumably associated with massive haloes.
Since the \Lya\ escape fraction is $\la10$\% from an emitting region of size
$\sim 1$ kpc, these objects will be difficult to find and hard to
detect, requiring $\sim 30$ -- 100 hours at each telescope pointing on
8 -- 10 m telescopes.  For CSR sources, existing large-format IR
arrays are close to ideal in terms of their noise characteristics for
conducting {\it wide-field} narrowband surveys where pixel sizes are
0.1\arcsec\ or larger.

However, we stress that \Lya\ can also arise from external large-scale
shocks (ELS) due to starburst winds, powered by CSRs, ploughing into
gas actively accreting onto the dark halo.  The winds effectively carry
energy from the dense, dusty environment of a starburst into lower
density regions, where the escape probability for \Lya\ photons is
greater.  ELS emission is expected to be considerably more clumpy ($\la
100$ pc) than CSR emission.  For ELS sources, IR arrays will need $1-2$
orders of magnitude improvement in dark current in order to detect
dispersed clumpy emission within the environments of massive haloes.
These sources require an IR camera with small pixels ($\sim
0.01$\arcsec) and adaptive optics correction (\eg\ GSAOI on Gemini),
and will therefore require a {\it targetted} observation of a Dark Age
source identified by a wide-field survey.  For either targetted or
wide-field surveys, the deepest observations will be those where the
pixel sampling is well matched to the size of the emitting regions.  We
show that there are only 3 -- 4 J-band windows [$z = 7.72, (8.22), 8.79,
9.37$] suitable for observing the Dark Ages in \Lya; we summarize their
cosmological properties in Table 1.

\end{abstract}


\section{Introduction}

In recent years, we have come to recognize that the `ionization epoch'
may lie just beyond our current observational horizon
\citep[\cf][]{fwd00,fss03,kes04}.  Structure formation models in CDM
cosmologies appear to show that this epoch took place at $z=7$ -- 12
\citep{go97,hl98}. Recent wide-angle
polarization measurements with WMAP \citep{ksb03} suggest the
ionization epoch could have been well under way by $z\sim 17$. However,
this may conflict with the inferred high column densities in $z\sim 6$
quasars \citep[\eg][]{wl04}.  Since cosmic ionization requires
only a tiny fraction of the primordial gas to be converted into stars
or black holes \citep{lb01}, it is possible to construct a
wide range of scenarios \citep{bcl99,bcl02,nu01,abn00,abn02,bl04,bl00,
  wl03,htl96,fwh01}.  Clearly, the first observations of the Dark Ages
will have a dramatic impact on our understanding of this new frontier.

In conventional CDM simulations, CSR sources are associated with the cores
of massive haloes. Since the distribution of dense peaks collapsing out
of an evolving Gaussian density field is well constrained
\citep{me03,p99}, bright CSR sources are expected to be rare. 
These will be difficult to find even with well optimized
wide-field surveys which exploit large pixels ($\sim$0.1\arcsec).
However, the number of detectable sources above a given flux level at a
fixed epoch is more uncertain. This requires a detailed understanding of
how and when \Lya\ emission is produced, and how it manages to escape its
immediate environment \citep{n91,hs99}.  Even the
most optimistic calculations \citep{bw87} show that \Lya\
detections will be a major observational challenge on 8 -- 10m telescopes.
The possible discovery of a lensed candidate galaxy at $z = 10$ by \citet{psr04}
demonstrates the power of lensing to extend our
observational reach.  However, the total number of sources accessible
this way is limited by the small total volume of the universe that is
strongly magnified by foreground lenses.

During the Dark Ages, gas accretion onto protogalactic cores must have
been well under way.  Galactic nuclei at the highest redshifts
observed to date ($z\sim5$) exhibit solar metallicities, and therefore
appear to have undergone many cycles of star formation \citep{hf99}.
Most galaxy cores early in their evolution must have 
experienced starburst-driven winds. Early protogalactic winds will have
carried large amounts of energy, with relatively low dust/metal content, 
away from the complex circumnuclear environment \citep{dnh04}.

In order to understand physical processes associated with the first
sources, we will need to resolve the \Lya\ structures. But, as we now
show, detecting emission powered by outflows is a different
experimental challenge from that posed by wide-field searches.

\bigskip
\bigskip
\section{The \Lya\ Challenge}

\subsection{Wind-scattered \Lya\ emission}

The difficulty of detecting \Lya\ during the Dark Ages is emphasized in
recent simulations by the GALFORM consortium \citep{l04,clb00}. 
The GALFORM simulations assume that all \Lya\ is
produced by star formation, and that 10\% escapes through the action of
galactic winds. The simulations are adjusted so as to reproduce the
Lyman-break and submillimetre galaxy number counts at presently
observed wavelengths. Down to a limiting flux magnitude of $f_{\rm lim}$
$=$ $3 \times 10^{-19}$ erg cm$^{-2}$ s$^{-1}$, they predict only about
ten sources per square arcmin will be observable across the {\it
entire} redshift range corresponding to the J band. The flux limit is
equivalent to a star formation rate of a few solar masses per year
which is easily high enough to drive large-scale winds \citep{h03,v03}.

The most efficient mechanism for \Lya\ escape is scattering by neutral
H entrained in an outflowing wind \citep{cn94}.  The best
observed starburst galaxy M82 reveals UV-scattered bicones
along the minor axis \citep{crb90,bgh90} on a
scale of 500 pc to 1 kpc. This corresponds to a spatial scale of
0.1 -- 0.2\arcsec\ at $z\sim 8$. Thus wide-field searches of rare
massive haloes will need to exploit pixels on this scale in order to
find CSR sources. The scattering wind may be clumped but, as we discuss 
below, the individual clumps are likely to fall below the detection
limit.

\subsection{Wind-driven \Lya\ emission}

We now examine the likelihood that most of the \Lya\ is distributed in
small clumps which are cooling out of dissipating gas, and spread over
a larger volume than \Lya\ arising directly from star forming regions.

During the early stages of galaxy formation, collapse is likely to be
comparable to star formation as a source of energy for the gas.  Cold
gas accumulated in the collapse of a small halo \citep{ro77} produces
an initial burst of star formation.  If the energy released into the
nascent interstellar medium by star formation greatly exceeded that
released in the collapse, then the bulk of the gas would be unbound
from the young galaxy.  On the other hand, the mechanical energy
required to limit the initial burst of star formation is comparable to
the binding energy of the gas, which is roughly the energy needed to
significantly rearrange gas within a galaxy.  Thus, if the burst of
star formation is self-limiting, its feedback cannot fall well short
of the binding energy of the gas, \ie\ the energy released by the
collapse.

Since radiative cooling times of gas at the virial temperature are much
shorter than dynamical times in low mass protogalaxies, feedback can
only be stored briefly as thermal energy.  Energy rapidly lost to
radiation is ineffective as feedback.  Only feedback energy that is
converted to kinetic (and later potential) energy can be effective in
limiting star formation on the dynamical timescale.  Effective feedback
needs to induce rapid large-scale flows, \ie winds, so we assume that
the primary channel of feedback is through winds.  This argument
requires a moderately high efficiency for the conversion of energy
released by star formation into wind energy, but that is consistent
with observations of starburst winds \citep[\eg][]{ss00}.

If collapse and the initial burst of star formation make comparable
energy inputs to the ISM, we can estimate both using the spherical
collapse model.  To form a disk, gas must dissipate at least the
vertical component of its velocity dispersion, \ie\ $\sigma^2/2$ per
unit mass.  For a dissipative collapse --- one producing cold gas --- the
energy dissipated is several times this, so we use the estimate of
$\sigma^2$ per unit mass dissipated by gas in the collapse.  The time
taken for a halo to virialize is roughly equal to its turn around time,
\ie\ half of its collapse time, $t_{\rm c}$.  If the energy dissipated
in the collapse is dissipated in half of the virialization time, the
dissipation rate $4 M_{\rm g} \sigma^2 / t_c$, where $M_{\rm g}$ is the
gas mass involved in the collapse.

Dark energy is insignificant for collapse at $z \sim 8$, so that the mean
density of a halo collapsing at $t_{\rm c}$ is $3 \pi / (G t_{\rm
c}^2)$, giving the virial radius for a halo of mass $M$ as $R = [G M
t_{\rm c}^2 /(4\pi^2)]^{1/3}$.  Treating the halo as an isothermal
sphere ($GM/R = 2\sigma^2$), then gives $\sigma^2 = 0.5 (2 \pi G M /
t_{\rm c})^{2/3}$, so that $\sigma \simeq 53 M_{10}^{1/3}\kms$, where
the halo mass $M = 10^{10} M_{10}\rm\ M_\odot$ and $t_{\rm c} =
6.27\times10^8$ yr [z = 8, in $\Lambda$CDM with $(h,\Omega_{\rm m},
\Omega_\Lambda) = (0.7, 0.3, 0.7)$].  Little \Lya\ can be produced in
the collapse unless the 
gas is heated over $\sim 10^4$ K, \ie unless $\sigma \gtrsim
10\kms$, requiring halo masses exceeding $10^8\rm\ M_\odot$.

Using the baryon fraction determined by WMAP \citep[$f_{\rm b} =
\Omega_{\rm b} / \Omega_{\rm m} \simeq 0.17$;][]{svp03} and assuming that
all baryons are gaseous in the collapse, the dissipation rate in the
gas during a collapse is $P_{\rm d} \simeq 4 f_{\rm b} M \sigma^2 /
t_{\rm c} \simeq 2 \times 10^{40} M_{10}^{5/3}\rm\ erg\ s^{-1}$.
Instantaneous dissipation rates will show a significant spread about
this value.  The
fraction emerging as \Lya\ photons depends on the shock temperature
and the fraction of photons that escape.  Higher escape fractions are
favoured by low optical depth and low abundances (low dust content).
In standard $\Lambda$CDM, very few $10^{10}$ M$_\odot$ halos collapse 
before z = 8 \citep{clb00}.

\begin{figure}
\psfig{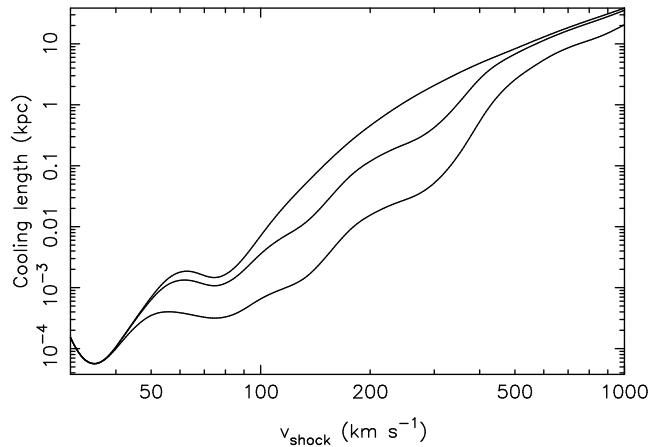}
\caption{
Cooling width of a collapse shock at $z=8$.  This shows the product of 
postshock cooling time and postshock speed as a function of the shock
speed, for gas with a preshock electron density of $0.036\rm\ cm^{-3}$ and
abundances of 0, 0.1 and 1 solar (top to bottom).  The cooling width
is inversely proportional to preshock density.} \label{cwidth}
\end{figure}

While the fraction of \Lya\ photons escaping from starbursts is small,
a significant part of the energy carried off by galactic winds is
ultimately likely to be dissipated in shocks.  These are another
potential source of \Lya\ photons and, from above, the energy in the
winds is comparable to that dissipated in the collapse.  The winds
effectively carry energy from the dense, dusty environment of a
starburst into lower density regions, where the escape probability for
\Lya\ photons is greater.  Starburst wind speeds are high and not
strongly dependent on the properties of the hosting halo.  This means
that if starbursts are triggered in halos smaller than $10^8\rm\
M_\odot$, then the terminating shocks of galactic winds may produce \Lya\
although the collapse shocks cannot.

From above, the mean baryon density in a dissipationless halo
collapsing at $z = 8$ is $3 \pi f_{\rm b}/ (G m_{\rm H} t_{\rm c}^2)
\simeq 0.036 \rm\ cm^{-3}$, where $m_{\rm H}$ is the mass of hydrogen.
Since dissipation significantly increases the density of the gas, the
typical density of gas running into shocks during a dissipative
collapse can be somewhat larger than this.  The collapse is fairly
chaotic and collapsing gas is likely to encounter several shocks
before finally reaching its destination.  Provided that the effects of
radiation transfer are
not too significant, the depth of the emitting region behind a
radiative shock is approximately equal to the product of the postshock
cooling time and velocity.  In a strong shock, the gas density is
increased by a factor of 4, the speed is decreased by the same factor
and the postshock temperature is $T_{\rm ps} = 3 \mu m_{\rm H} v_{\rm
s}^2 / (16 k)$, where $v_{\rm s}$ is the shock speed and $\mu m_{\rm
H}$ is the mean mass per particle.  Fig.~1 shows the width of the
postshock cooling region for a preshock electron density of $0.036\rm\
cm^{-3}$, for metal abundances of 0, 0.1 and 1 solar.

Initially, the collapsing gas flow may be coherent on larger scales
than the postshock cooling length in Fig.~1.  In that case, cooling
shocked regions are sheetlike and emission is brightest at caustics
where we see folds in these sheets projected onto the sky.  The cooling
gas is subject to thermal and other instabilities that will generally
cause it to fragment on about the scale of the cooling length.  (The
flow into any further shocks is likely to be considerably more
chaotic.) For haloes in the mass range of interest, the shock speeds in the
collapse are, at most, 100 -- 200 $\rm\ km\ s^{-1}$. In a pristine gas,
the inferred size of cooling clumps is roughly 100~pc, or
0.02\arcsec\ at a redshift of $z\sim 8$.  The small size of the cooling
region demands an infrared imager which utilizes small pixels ($\sim
0.01$\arcsec) and adaptive optics correction.  However, we point out
that rapid winds from starbursts can produce significantly larger
cooling regions.

Since gas is compressed significantly in a dissipative collapse, the
time scale for star formation in a clump of collapsed gas can be much
shorter than the dynamical time of the collapsing system ($\sim 10^8$
yr).  Massive stars take only $\sim 10^6$ yr to produce supernovae,
giving plenty of time for starbursts and their winds to get underway
while other gas continues to collapse into the system.  Interaction
between infalling gas and starburst winds can lead to further
shocking of both.  This process is observed in M82 where the outflowing
starburst-driven wind impinges directly on infalling gas at a radius of
11~kpc \citep{yhl94} producing observable \Ha\ and x-ray emission 
\citep{db99,lhw99}.

\begin{figure}
\psfig{file=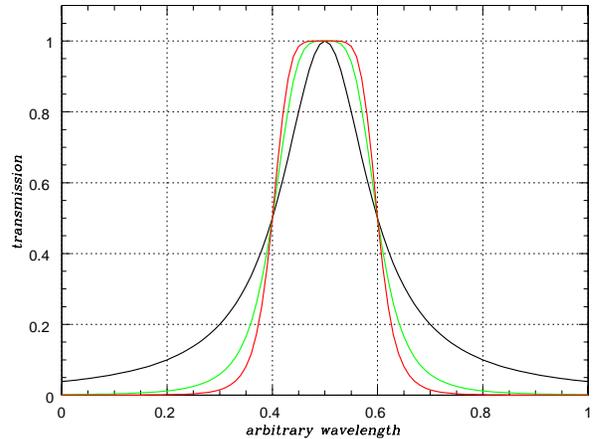,width=8.5cm,angle=0}
\caption{
Filter profiles used in our GSAOI calculations. The Lorentzian
profile ($m=1$) shown in black is the expected response of a high finesse
(N=40) tunable filter. The green profile ($m=2$) is the low-cost 
filter option, the blue profile ($m=3$) is the high-cost filter
option.} \label{butterworth}
\end{figure}

\section{Expected \Lya\ surface brightness}

We now examine whether wind-powered \Lya\ emission in small clumps can
be detected with an 8m telescope using an adaptive optics imager with
small pixels (0.02"). We assume that the total energy released in
\Lya\ at a fixed star formation rate is comparable to what is generated
in large-scale shocks driven by the starburst, and that this emission
escapes without attenuation.  As discussed in \S 2.2, this assumes a
high conversion rate of the supernova energy into wind energy
\citep[\eg][]{ss00}.

From surveys of nearby dwarf galaxies \citep{m03,v03}, we
adopt a wind radius of 5~kpc. At $z\sim 8$, the shock occurs at 1\arcsec\
radius and is barely resolved. Naively, if the wind energy escapes along
bipolar cones, the \Lya\ surface brightness will be increased by the
decrease in solid angle compared to a spherical wind.  This factor can be
$\sim 10$ since most winds appear to be highly collimated
\citep{sb98,sdw03}.  However, this
factor does not apply to a comparison with the expected flux from a CSR,
since the normal assumption is that the collimated wind material is what
renders the central source visible. An escape fraction of 10\% assumed
in the GALFORM simulations is the fraction of radiation which escapes
along the bicones multiplied by the fraction scattered in the wind.

For wind-induced emission, in our model, the total power available is
$10\:f_{\rm lim}$ but this is now spread over many more pixels.  The 
surface brightness of the shock induced \Lya\ emission depends
critically on how the bipolar shock surfaces are oriented with respect
to the observer.  If the bipolar wind lies in the plane of the sky,
most of the emission is confined to limb-brightened arcs over an area
of $\sim 10^2$ pixels depending on the curvature of the shock surface.
If the bipolar wind is directed at the observer, the projected radius
of the shocked-induced nebula is half the intrinsic radius ($60\degr$
opening angle). Thus, the flux is now dispersed over an area of $\sim
10^3$ pixels.

Our expectation is that the \Lya\ flux {\it per pixel} will be an order
of magnitude fainter in external large-scale shocks (ELS) compared to
CSR emission in a large pixel survey (0.1"). We believe that it will be
important to reach this flux level if we are to understand the nature
and origin of the \Lya\ emission.  We have assumed that essentially all
of the \Lya\ emission escapes its environs since it is produced {\it in
situ}.  But in the presence of any diffuse intergalactic HI component
at rest with the expanding universe, the shock surface must have a
sufficient redshift in order for the \Lya\ emission to escape to the
observer.

We have already noted that CSR emission is rendered visible through
scattering by neutral H in the wind. From the local universe, we
know that winds arise from highly inhomogeneous starburst cores and
entrain most of their mass from cooler material in the surrounding disk
\citep{sbh96}.  Thus the scattering medium is likely to be highly
clumped. However, since the total escaping \Lya\ flux in our model is 
$f_{lim}$, these scattering clumps are likely
to be an order of magnitude fainter than self-illuminating clumps of
shocked gas induced by the same wind.

\begin{deluxetable}{@{}lllll}
\tablewidth{\linewidth}
\tablecaption{Basic cosmological parameters for the four dark J-band
windows. \label{cospars}}
\startdata
Wavelength ($\mu$m)   & 1.06 & 1.12 & 1.19 & 1.26 \\
Redshift    & 7.72 & 8.22 & 8.79 & 9.37 \\
Time since Big Bang (Gyr)  & 0.66 & 0.60 & 0.55 & 0.51 \\
Time before present (Gyr)  & 12.80 & 12.85 & 12.91 & 12.95 \\
Physical scale (kpc/arcsec)  & 4.92 & 4.73 & 4.53 & 4.35 \\
Luminosity distance, $d_L$ (Mpc)  & 77283 & 82979 & 89667 & 96398 \\
Flux, $F=L_S/4\pi d_L^2\ (10^{-19}\rm cgs)$\tablenotemark{a} & 1.40 & 1.21 & 1.04 & 0.90 \\
Physical depth, D (Mpc/1000 km s$^{-1}$) & 1.009 & 0.929 & 0.849 & 0.779 \\
Comoving depth, D$_o$ (Mpc/1000 km s$^{-1}$) & 8.798 & 8.565 & 8.312 & 8.078 \\
Physical volume, 1'$\times$1'$\times$D (Mpc$^3$) & 0.088 & 0.075 & 0.063 & 0.053 \\
Comoving volume, 1'$\times$1'$\times$D$_o$ (Mpc$^3$) & 58.5 & 58.7 & 59.0 & 59.1 \\
\enddata
\tablecomments{The cosmology is $(h, \Omega_{\rm m}, \Omega_\Lambda) = (0.7, 0.3,
0.7)$.  The comoving volume is essentially constant over
the windows and the total timespan is only 150 Myr.}
\tablenotetext{a}{The expected flux from a source with luminosity $L_S
  = 10^{41}$ erg s$^{-1}$, in units of 10$^{-19}$ erg cm$^{-2}$
  s$^{-1}$, only varies by 50\%.} 
\end{deluxetable}

\section{Survey Method}

\subsection{Wide-field \Lya\ survey}

We envisage an initial survey with a wide-field, large-pixel imager
which scans over several fields in order to identify an initial list of
CSR sources.  This is the primary motivation of the DAZLE instrument
under construction at the Institute of Astronomy, University of
Cambridge for the Very Large Telescope \citep[see also
http://www.ast.cam.ac.uk/\~ optics/dazle]{b03} which exploits high performance
$\Res=1000$ interference filters closely spaced in wavelength within
the J band. The wide-field (6.9\arcmin\ square) observations are built up in
interleaved sub-exposures by switching between the two filters: the
images are then differenced in order to detect signals in one band that
are not evident in the other.  This technique has been widely used with
the Taurus Tunable Filter (TTF) on the AAT and has led to the
identification of a line-emitting populations out to $z \sim 5$
\citep[\eg][]{bbb04} .

Each target field with DAZLE will require long exposures, and may
require several fields in order to find a single source.  Table~1
summarises the cosmological properties of the four `dark' windows in
the J-band originally identified in the DAZLE design study \citep[see
also http://www.aao.gov.au/dazle]{c03} which are clearly evident 
in Fig.~3.

We now present SNR calculations for a wide-field and an adaptive optics 
near-IR imager on an 8m
telescope. The calculations utilize both airglow and absorption spectra
at a spectral resolution of $\Res = 10\:000$ \citep[\cf][]{ob98}.  We
adopt instrumental parameters for DAZLE and for the Gemini
South Adaptive Optics Imager (GSAOI) to be commissioned in 2005: the
key numbers are listed in Table~2 \citep{mcg03}.  Note that
several of the parameters are expressed over a range: the lower value
is used in our ELS calculation (GSAOI), and the upper value is used in
our CSR calculation (DAZLE). The GSAOI will utilise a HAWAII-2RG array,
under development for the James Webb Space Telescope, which is expected
to have exceptionally low dark current. DAZLE exploits a HAWAII-2 array
which has a much higher dark current, but which is adequate for wide-field
surveys, as we show below.

\begin{deluxetable}{lllll}
\tablewidth{\linewidth}
\tablecaption{Basic parameters for GSAOI and a DAZLE-like instrument
operating at the Gemini telescope.}
\startdata
Reduced telescope area & $(\pi/4)(7902$ -- $1302)^2\rm\ cm^{-2}$ \\
System throughput & 0.27 \\
Filter throughput & 0.80 \\
Detector pixel size & 0.02" -- 0.1" \\
Detector read noise & $5\rm\ e^-\ pix^{-1}$ \\
Detector dark current & 0.003 -- $0.05\rm\ e^-\ s^{-1}\ pix^{-1}$ \\
Night-glow continuum & 2 -- 5 Rayleigh \AA$^{-1}$ \\
Source flux & 3 -- $30 \times 10^{-20}\rm\ erg\ cm^{-2}\ s^{-1}$ \\
Source diameter & 0.04" -- 0.20" \\
Total SNR for source & 3 -- 5 \\
Exposure time & 3600 s \\
\enddata
\tablecomments{For parameters shown over a range, the lower bound is
specific to the ELS (GSAOI) case, the upper bound refers to the CSR
(DAZLE) case.  The one exception is the night sky continuum where we
incorporate both bounds in both calculations.
The reduced telescope area incorporates the loss due
to the Cassegrain hole (M2 stop). The filter throughput assumes an
IR detector that cuts off at 2.5$\mu$m;  this reduces to 60\% for a detector
cut-off at 5$\mu$m due to the need for additional optical density layers
for long wavelength suppression.  The exposure time determines the number
of read noise contributions to the summed image.}
\end{deluxetable}

\begin{figure}
\psfig{file=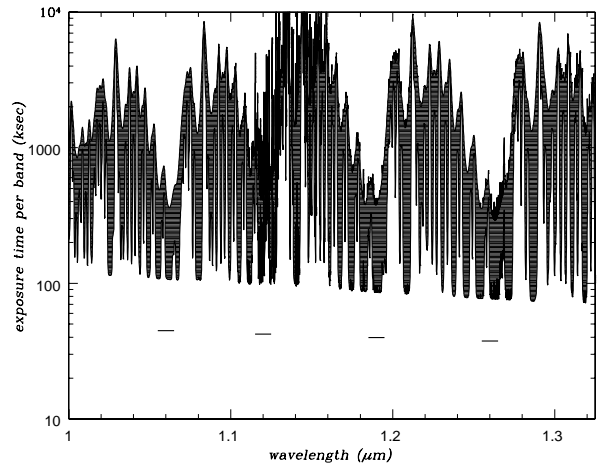,width=8.5cm,angle=0}
\caption{
The calculated total exposure times (in ksec) for the parameters
listed in Table 2. The solid horizontal lines are the times to reach
SNR $=$ 5 in an initial DAZLE survey. The filled region shows the 
expected times to reach SNR $=$ 3 in a targetted study with GSAOI:
the lower envelope is for a filter with $m=3$, \Res=1000; the upper envelope 
is for $m=1$, \Res=500.  The dark windows near 1.06$\mu$m,
1.19$\mu$m and 1.26$\mu$m are the most suitable for \Lya\ 
observations; note how the upper envelope picks these out. The 
cosmological parameters for these four windows are given in Table 1.}
\label{gsaoi}
\end{figure}

\smallskip
The equivalent luminosity for the quoted \Lya\ flux and the physical
size of the \Lya\ blobs is given in Table~1 as a function of redshift.
Since the blob sizes in our calculations are larger than the
AO-corrected psf of GSAOI, we do not consider the J-band Strehl ratio
in our calculations.  The J-band airglow spectrum was normalized to the
expected J-band counts given in \citet{mcg03}.  We consider two
limiting cases for the night-glow continuum between the OH lines
\citep[\eg][]{c96}; the actual night-glow continuum level is unknown 
but is bracketed by the quoted values in Table 2.
The night-glow surface brightness is quoted in Rayleighs (R) per Angstrom
where 1 R $=$ 10$^6$/$4\pi$ phot cm$^{-2}$ s$^{-1}$ sr$^{-1}$.

At a fixed spectral bandpass defined by \Res, the wings of the filter
profile must be considered.  \citet{jbb96} demonstrate
that the out-of-band blocking of an interference filter with $m$ cavities
is closely matched to a Butterworth function of degree $m$.  We have
incorporated the Butterworth profile (see Fig.~2) in our calculations.
The single cavity ($m=1$) is the Lorentzian profile of a tunable filter
in the high finesse limit \citep{bj98}. The relatively
low-cost DAZLE filters utilise $m=2$ (two cavity), although there is
also a high-cost option with $m=3$. We assume that placement within the
instrument does not degrade the effective bandpass, an issue discussed
at length by Bland-Hawthorn\etal\ (2001).

Both $m=2$ and $m=3$ filters are risky and highly expensive items to
manufacture since the multi-cavity dielectric coatings can exceed 10$\mu$m
in total thickness, requiring hundreds or even thousands of layers.
As was found in the DAZLE study, this can be greatly exacerbated by the
need for high optical density to achieve long-pass blocking\footnote{An
alternative strategy is to exploit an IR array with a sensitivity cut-off 
at a shorter wavelength ($\lambda<2\mu$m).  If the dark current can be 
kept to a minimum, we see this approach as preferable to a HAWAII-2RG array.} 
by the interference filter.

In Fig. 3, the horizontal lines indicate we need 40~ksec per field in
order to achieve a 5$\sigma$ detection in one or other band, but note
that the DAZLE technique surveys twice the volume of a single
$\Res=1000$ filter image. (If the detection relies on the differencing
of on-line and off-line bands, its statistical significance is reduced
to 3.5$\sigma$.) An (almost) equivalent strategy is to halve the
exposures per field, and to observe two fields in two closely spaced
bands.  Once tentative sources have been identified, in order for the
emission to be \Lya, the source should not be evident in deep $ugriz$
images (\ie\ AB mag $\ga$ 28) corresponding to rest-frame Lyman
continuum.

\subsection{Targetted \Lya\ survey}

The initial wide-field survey will do little more than identify
tentative sources for closer study. At this point, we target specific
sources for further study with a high resolution imager and adaptive
optics.  The initial survey will need to target fields in the 
vicinity of IR-bright stars which can be utilised for AO correction.
A particular advantage of narrowband filters is that the AO
correction is not hampered by atmospheric refraction which can spread
point sources over tens of pixels within a broadband filter.

For a targetted study, we require only a single narrowband filter, and
need only achieve $3\sigma$ per blob within the \Lya\ nebula to map the
distribution of ionized gas. But Fig. 3 shows that the total exposure
times per source are 100 ksec.  At $\Res=1000$, there appear to be a
dozen useful windows within the J-band. However, in our design study
for DAZLE \citep{c03}, we find that only $3-4$ windows are practical.
For a filter placed at the pupil, there is a slight phase effect over
the field such that the passband at the filter edge is bluer than at
the centre of the field. For a filter placed in the converging beam,
the passband is broadened slightly.

Another concern is the difficulty of manufacturing a 0.1\%$\lambda$
narrowband filter centred at the prescribed wavelength. Given the
restricted tuning capability of an interference filter, and its degraded
performance on tilting, it is tempting to consider a tunable filter for
\Lya\ imaging. However, the upper envelope of the filled region in Fig.~3
shows the expected poor performance of a filter profile with Lorentzian wings.

The above considerations require that we accept a window which is
actually twice as broad as the design bandwidth.  In Fig.~3, we note
that there is only a handful of good windows available at $\Res =
500$. In terms of photometric stability, the windows near 1.06$\mu$m,
1.19$\mu$m and 1.26$\mu$m are the most ideal.  The window at 1.12$\mu$m
suffers from variable and complex atmospheric absorption.  The relevant
cosmological parameters for these four windows are given in Table~1.

The 1.12$\mu$m window could be recovered if two IR arrays were placed
side by side in the image plane, and these devices were moved back and
forth every few mins, in synchrony with nodding by the telescope, such
that we build up two separate fields of view. This mechanical form of
`nod \& shuffle' \citep[\cf][]{gb01} would average
out atmospheric effects, and at the same time minimize the read noise
penalty, although it incurs a $\sqrt{2}$ penalty from dark current noise.

If systematics can be controlled, the long total exposure times of 30
hr per field are not unreasonable.  If there are weak OH features in
the dark windows, these are likely to be variable and may define the
systematic noise limit. Note that if pixel stability dominates the
noise limit, a Dark Age experiment cannot be carried out since this
leads to short exposures (Fowler-8 sampling) and results in a huge
read noise penalty. However, D.~B.~Hall (2003, personal communication)
has found that the HAWAII-2RG arrays have comparable stability to some
of the best optical detectors.

\section{The Importance of Dark Current}

We now show that the IR arrays specified in Table~2 are ideally suited
to the proposed experiments.  If the J-band windows are truly dark, the
read $+$ dark noise contribution defines the limiting sensitivity we
can reach. This is illustrated in Fig.~4 which shows the total exposure
time required in the 1.06$\mu$m window to achieve SNR $=$ 3 on a \Lya\
blob with a total flux of $3\times 10^{-20}$ erg cm$^{-2}$ s$^{-1}$.
Here we assume a filter with $\Res=1000$ and $m=3$.  This calculation is
specific to GSAOI:  the pixels are 0.02" and the read noise is 5 e$^-$. At
the top of the figure, the blobs are spread over many pixels which
greatly increases the effective background from dark current.

The results are only weakly sensitive to the number of separate exposures
since the dark current noise dominates for exposures longer than about 500
sec. Note in particular how a targetted study with GSAOI benefits from
a very low dark current array (d.c. $<$ 0.01 e$^-$ s$^{-1}$ pix$^{-1}$)
for essentially any blob size. The HAWAII-2RG array has demonstrably
the lowest dark current of any existing or planned array technology 
(D.B. Hall 2003, personal communication).

In Fig.~5, we show the total exposure time required in the 1.06$\mu$m
window to achieve SNR $=$ 5 on a \Lya\ blob with a total flux of
$3\times 10^{-19}$ erg cm$^{-2}$ s$^{-1}$.  This calculation is
specific to the initial DAZLE survey with $\Res=1000$ and $m=3$. Note
how the uncertainty in the dark sky continuum leads to some uncertainty
in the detectability of \Lya\ blobs. For CSR source sizes of 0.2", the
existing HAWAII-2 arrays are well suited to wide-field \Lya\ surveys of
this kind.

The implication here is that the pixel size must be well matched to the
expected size of the \Lya\ emitting regions. If the pixels are much
larger than the emitting blobs, night-glow continuum will dominate the
exposure and result in excessively long exposures. If the emitting
blobs are much larger than the pixels, dark current dominates leading
to excessively long exposures.

\begin{figure}
\psfig{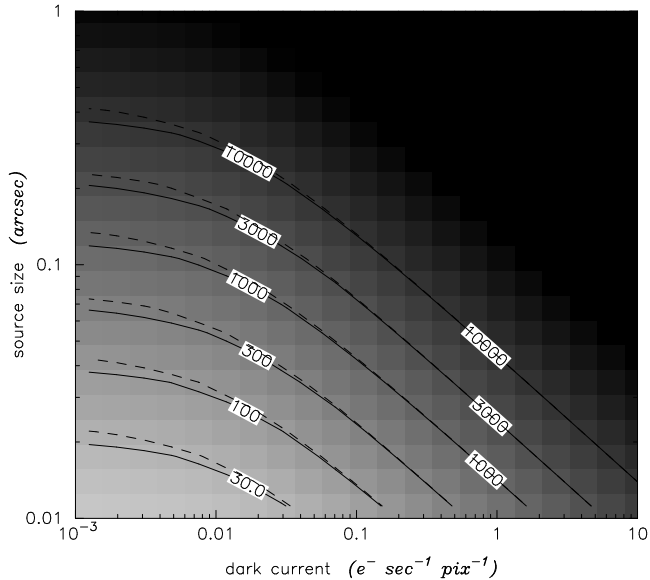}
\caption{
Total exposure times (in ksec) in the 1.06$\mu$m window as a function
of dark current (horizontal) and blob size (vertical; in arcsec)
for a detector with 0.02\arcsec\ pixels. We use the parameters in Table 2
specific to GSAOI. The exposure times assume SNR=3 per source for a total
object flux of $3\times 10^{-20}$ erg cm$^{-2}$ s$^{-1}$.  Note the 
importance of very low dark current in order to detect clumpy \Lya\ emission.
The solid lines assume a high level of night-glow continuum between the OH lines
(see Table 2), the dashed lines assume weak night-glow. } 
\label{gsadc}
\end{figure}

\begin{figure}
\psfig{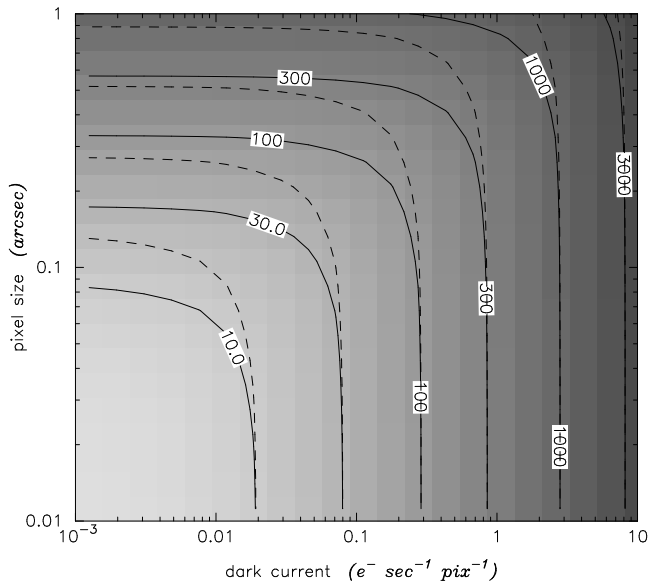}
\caption{
Total exposure times (in ksec) in the 1.06$\mu$m window as a function
of dark current (horizontal) and pixel size (vertical; in arcsec). We
use the parameters in Table 2 specific to a DAZLE-like instrument. 
The exposure times assume SNR=5 per source for a total
object flux of $3\times 10^{-19}$ erg cm$^{-2}$ s$^{-1}$.
The solid lines assume a high level of night-glow continuum between the OH lines
(see Table 2), the dashed lines assume weak night-glow. For 0.2" source
diameters, existing HAWAII-2 arrays are well suited to wide-field
searches.} 
\label{gsadd}
\end{figure}

\bigskip
\section{Concluding remarks}

It is clear that something quite extraordinary took
place over a time span of a few hundred million years which led to 
the complete ionization of the intergalactic medium.  
The epoch of (re)ionization is likely to attract a great deal 
of attention over the coming decade.

We have shown that detecting sources beyond $z=7$
is going to be very challenging on 8 -- 10m telescopes. 
But since 30 -- 100m telescopes will not be in operation
until the next decade, narrowband imaging is likely to dominate
studies of the Dark Ages for years to come.
We have outlined a strategy based on wide-field 
surveys to identify star formation in massive haloes, followed
by detailed studies with an adaptive optics imager. 
The optimal technologies for both kinds of study will be
available in the next few years.

It is often observed that major discoveries are made within 
five years of a new technology \citep[\eg][]{h81}.  What is less
well known is that this statement normally applies to
experiments which achieve the systematic noise limit of
the apparatus.  \citet{gb01} argue that
the systematic limit possible with nod \& shuffle spectroscopic
observations argues for total integration times measured in
months or even years.  Since large telescopes are general 
user facilities, it is relatively rare that hundreds of hours
are devoted to a single target field. But in order to reach
back to the Dark Ages, observing programmes of one or two weeks at a 
time will be essential.

\acknowledgments
PEJN was partly supported by NASA grant NAS8-01130.  JBH would like to
thank Warrick Couch and Doug Simons for the impetus to think about
Dark Age science in the context of Gemini, and to the GSAOI team at
Mount Stromlo (PI: Peter McGregor) for their excellent work.


\begin{thebibliography}{}

\bibitem[Abel, Bryan \& Norman(2000)]{abn00}
Abel, T., Bryan, G. L., \& Norman, M. L. 2000, ApJ, 540, 39
\bibitem[Abel, Bryan \& Norman(2002)]{abn02}
Abel, T., Bryan, G. L., \& Norman, M. L. 2002, Science, 295, 93
\bibitem[Barkana \& Loeb(2000)]{bl00}
Barkana, R., \& Loeb, A. 2000, ApJ, 539, 20
\bibitem[Baron \& White(1987)]{bw87}
Baron, E., \& White, S. D. M. 1987, ApJ, 322, 585
\bibitem[Barr \etal(2004)]{bbb04}
Barr, J. M., \etal\ 2004, submitted to MNRAS
\bibitem[Bland-Hawthorn \& Jones(1998)]{bj98}
Bland-Hawthorn, J. \& Jones, D. H. 1998, PASA, 15, 44
\bibitem[Bland-Hawthorn \etal(2001)]{bvg01}
Bland-Hawthorn, J., van Breugel, W., Gillingham, P. R., Baldry, I. K.,
\& Jones, D. H. 2001, ApJ, 563, 611
\bibitem[Bland-Hawthorn \etal(2003)]{b03}
Bland-Hawthorn, J. 2003, AAO Newsletter, 103, 16
\bibitem[Blecha \etal(1990)]{bgh90}
Blecha, A., Golay, M., Huguenin, D., Reichen, D, \& Bersier, D. 1990,
A\&A, 233, L9 
\bibitem[Bromm, Coppi \& Larson(1999)]{bcl99}
Bromm, V., Coppi, P. S., \& Larson, R. B. 1999, ApJ, 527, L5
\bibitem[Bromm, Coppi \& Larson(2002)]{bcl02}
Bromm, V., Coppi, P. S., \& Larson, R. B. 2002, ApJ, 564, 23
\bibitem[Bromm \& Larson(2004)]{bl04}
Bromm, V. \& Larson, R. B. 2004, ARAA, in presss
\bibitem[Chen \& Neufeld(1994)]{cn94}
Chen, W. \& Neufeld, D. A. 1994, ApJ, 432, 567
\bibitem[Cianci(2003)]{c03} Cianci, S. 2003, PhD, University of Sydney
\bibitem[Cole \etal(2000)]{clb00}
Cole, S., Lacey, C. G., Baugh, C. M., \& Frenk, C. S. 2000, MNRAS,
319, 168 
\bibitem[Content(1996)]{c96} Content, R. 1996, ApJ, 464, 412
\bibitem[Courvoisier \etal(1990)]{crb90}
Courvoisier, T. J.-L., Reichen, M., Blecha, A., Golay, M., \&
Huguenin, D. 1990, A\&A, 238, 63
\bibitem[Desjacques \etal(2004)]{dnh04}
Desjacques, V., Nusser, A., Haehnelt, M. G., \& Stoehr, F. 2004,
MNRAS, astro-ph/0311209
\bibitem[Devine \& Bally(1999)]{db99}
Devine, D., \& Bally, J. 1999, ApJ, 510, 197
\bibitem[Fan \etal(2000)]{fwd00} Fan, X. \etal\ 2000, AJ, 120, 1167
\bibitem[Fan \etal(2003)]{fss03} Fan, X. \etal\ 2003, AJ, 125, 1649
\bibitem[Fryer, Woosley \& Heger(2001)]{fwh01}
Fryer, C., Woosley, S. E., \& Heger, A. 2001, ApJ, 550, 372
\bibitem[Glazebrook \& Bland-Hawthorn(2001)]{gb01}
Glazebrook, K., \& Bland-Hawthorn, J. 2001, PASP, 113, 197
\bibitem[Gnedin \& Ostriker(1997)]{go97}
Gnedin, N. Y. \& Ostriker, J. P. 1997, 486, 581
\bibitem[Haiman \& Loeb(1998)]{hl98}
Haiman, Z. \& Loeb, A. 1998, ApJ, 503, 505
\bibitem[Haiman \& Spaans(1999)]{hs99}
Haiman, Z. \& Spaans, M. 1999, ApJ, 518, 138
\bibitem[Haiman, Thoul \& Loeb(1996)]{htl96}
Haiman, Z. Thoul, A. A. \& Loeb, A. 1996, ApJ, 464, 523
\bibitem[Hamann \& Ferland(1999)]{hf99}
Hamann, F. \& Ferland, G. 1999, ARAA, 37, 487
\bibitem[Harwit(1981)]{h81} 
Harwit, M. 1981, Cosmic Discovery, New York: Basic Books
\bibitem[Heckman(2003)]{h03}
Heckman, T.M. 2003, In Galaxy Evolution: Theory \& Observations, eds
V. Avila-Reese, C. Firmani, \& C. Allen, Rev. Mex. A. A., 17, 47
\bibitem[Jones, Bland-Hawthorn \& Burton(1996)]{jbb96}
Jones, D. H., Bland-Hawthorn, J., \& Burton, M. G. 1996, PASP, 108,
929
\bibitem[Kneib \etal(2004)]{kes04}
Kneib, J.-P., Ellis, R. S., Santos, M. R., \& Richard, J. 2004, ApJ,
astro-ph/0402319
\bibitem[Kogut \etal(2003)]{ksb03}
Kogut, A. \etal\ 2003, ApJS, 148, 161
\bibitem[Lacey \etal(2004)]{l04} Lacey, C. \etal\ 2004, in preparation
\bibitem[Lehnert, Heckman \& Weaver(1999)]{lhw99}
Lehnert, M., Heckman, T. M., \& Weaver, K. A. 1999, ApJ, 523, 575
\bibitem[Loeb \& Barkana(2001)]{lb01}
Loeb, A., \& Barkana, R. 2001, ARAA, 39, 19
\bibitem[McGregor \etal(2003)]{mcg03}
McGregor, P. \etal\ 2003, GSAOI Critical Design Review Vol. 2
(84699-GEM00334)
\bibitem[Martin(2003)]{m03}
Martin, C.L. 2003, In Galaxy Evolution: Theory \& Observations, eds
V. Avila-Reese, C. Firmani, \& C. Allen, Rev. Mex. A. A., 17, 56
\bibitem[Miralda-Escud\'e(2003)]{me03}
Miralda-Escud\'e 2003, Science, 300, 1904
\bibitem[Nakamura \& Umemura(2001)]{nu01}
Nakamura, F., \& Umemura, M. 2001, ApJ, 548, 19
\bibitem[Neufeld(1991)]{n91} Neufeld, D. A. 1991, ApJ, 370, L85
\bibitem[Offer \& Bland-Hawthorn(1998)]{ob98}
Offer, A. R. \& Bland-Hawthorn, J. 1998, MNRAS, 299, 176
\bibitem[Peacock(1999)]{p99} 
Peacock, J.A. 1999, Cosmological Physics, CUP
\bibitem[Pell\'o \etal(2004)]{psr04}
Pell\'o, R., Schaerer, D., Richard, J., Le Borgne, J.-F., \& Kneib,
J.-P. 2004, A\&A, 416, L35 (astro-ph/0403025)
\bibitem[Rees \& Ostriker(1977)]{ro77}
Rees, M. J., \& Ostriker, J. P. 1977, MNRAS, 179, 541
\bibitem[Shopbell \& Bland-Hawthorn(1998)]{sb98}
Shopbell, P. L., \& Bland-Hawthorn, J. 1998, ApJ, 493, 129
\bibitem[Spergel \etal(2003)]{svp03}
Spergel, D. N. \etal\ 2003, ApJS, 148, 175
\bibitem[Strickland \& Stevens(2000)]{ss00}
Strickland, D., \& Stevens, M. R. 2000, MNRAS, 314, 511
\bibitem[Suchkov \etal(1996)]{sbh96}
Suchkov, A. A., Berman, V. G., Heckman, T. M., \& Balsara, D. S. 1996,
ApJ, 463, 528
\bibitem[Sugai, Davies \& Ward(2003)]{sdw03}
Sugai H., Davies, R. I., \& Ward, M. J. 2003, ApJ, 584, L9
\bibitem[Veilleux(2003)]{v03}
Veilleux, S. 2003, In Recycling Intergalactic and Interstellar Matter,
IAU Symp. 217, 168
\bibitem[Wyithe \& Loeb(2003)]{wl03}
Wyithe, S., \& Loeb, A. 2003, ApJ, 588, L69
\bibitem[Wyithe \& Loeb(2004)]{wl04}
Wyithe, S., \& Loeb, A. 2004, Nat, in press
\bibitem[Yun, Ho \& Lo(1994)]{yhl94}
Yun, M. S., Ho, P. T. P., \& Lo, K. Y. 1994, Nat, 372, 530

\end{thebibliography}
\end{document}